\documentclass[12pt]{article}
\usepackage[scale={0.76,0.85}]{geometry}
\usepackage[latin1]{inputenc}
\usepackage{amsmath,amssymb,bbm,mathrsfs}
\usepackage[dvips]{epsfig}
\usepackage{axodraw,rotating,graphicx,subfigure,caption2}
\usepackage{hyperref}

\providecommand{\eprint}[2][]{\href{http://arxiv.org/abs/#2}{#2}}

\newcommand{\cc}{\text{\sc c}}
\newcommand{\ckm}{\text{\sc ckm}}

\newcommand{\eabg}{\epsilon_{\alpha\beta\gamma}}
\newcommand{\ew}{\text{\sc ew}}
\newcommand{\Le}{\text{\sc l}}
\newcommand{\GG}{\text{\sc gg}}
\newcommand{\gut}{\text{\sc gut}}
\newcommand{\hbcc}{\ol{H}_{\cc}}
\newcommand{\hbf}{\ol{H}_{\!f}}
\newcommand{\ps}{\text{\sc ps}}
\newcommand{\ri}{\text{\sc r}}
\newcommand{\sm}{\text{\sc sm}}

\DeclareMathOperator{\diag}{diag}
\DeclareMathOperator{\tr}{tr}

\newcommand{\braket}[3]{\left\langle #1\!\left| #2\right|\! #3\right\rangle} 
\newcommand{\VEV}[1]{\left\langle #1\right\rangle}        
\newcommand{\SLASH}[2][3]{\ensuremath{\rlap{\raisebox{1pt}{$\mskip #1 mu /$}}#2}}

\newlength{\textlength}
\newlength{\overlinelength}

\newcommand{\ol}[2][.625]{
   \settowidth{\textlength}{$#2$}
   \setlength{\overlinelength}{3pt}
   \addtolength{\overlinelength}{0.4\textlength}
   \makebox[\textlength][s]{$#2$}
   \hspace{-.5\textlength}\hspace{-\overlinelength}\hspace{#1\overlinelength}
   \overline{
      \makebox[\overlinelength][s]{
         \vphantom{$#2$}
      }
   }
   \hspace{-#1\overlinelength}\hspace{.5\textlength}
}

\begin{document}


\title{ {\normalsize
    July 2004 \hfill DESY 04-125 \\[-3mm]
    \hspace*{\fill} hep-ph/0407173
    }\\
  \vspace{15mm} \textbf{Proton Decay in Supersymmetric GUT
    Models}
  \\[5mm]}

\author{S{\"o}ren~Wiesenfeldt\\
  {\normalsize \slshape Deutsches Elektronen-Synchrotron DESY,
    Hamburg, Germany} }

\date{}

\maketitle

\begin{abstract}
  \noindent
  The instability of protons is a crucial prediction of supersymmetric
  GUTs.  We review the decay in minimal supersymmetric {\sf SU(5)},
  which is dominated by dimension-five operators, and discuss the
  implications of the failure of Yukawa unification for the decay
  rate.  In a consistent {\sf SU(5)} model, where {\sf SU(5)}
  relations among Yukawa couplings hold, the proton decay rate can be
  several orders of magnitude smaller than the present experimental
  bound.  Finally, we discuss orbifold GUTs, where proton decay via
  dimension-five operators is absent.  The branching ratios of
  dimension-six decay can significantly differ from those in four
  dimensions.
  \\[3mm]
  {\footnotesize Keywords: baryon number violation; grand unified
    theory, orbifold; supersymmetry.
    }
  \\[5mm]
\end{abstract}



\noindent
The stability of protons has been a research topic for a long time
\cite{baryon-number}, and exactly 50 years ago, the first experiments
started to measure its lifetime \cite{exp} --- although no mechanism
was known that leads to proton decay.  The situation changed when both
baryon number violating processes were found in the Standard Model
(SM) \cite{'tHooft76} and the idea of Grand Unification came up, where
the SM is embedded in a simple gauge group \cite{georgi74}.  Today, it
is widely believed that protons will decay, even if they do so after a
tremendously long time.

Grand Unified Theories predict proton decay and the unsuccessful
search for decaying protons gives a strong constraint on GUT models.
The first GUT model, minimal {\sf SU(5)} \cite{georgi74}, could
already be excluded in the early 1980s.  Recently, it was claimed that
even its supersymmetric version \cite{susy-su5} is excluded by the
experimental limit on proton decay \cite{goto99,murayama02}.  In this
review, we discuss these analyses and point out the crucial
assumptions which have been made to analyze the dominant
dimension-five operators.  We will see that the problem of minimal
supersymmetric {\sf SU(5)} is less proton decay than the failure of
Yukawa unification.  Hence, we need a consistent model which explains
the Yukawa couplings and we will see that a supersymmetric {\sf SU(5)}
model with minimal particle content cannot be excluded by proton decay
\cite{cw03}.

Although such models are in agreement with the experimental limits on
proton decay, the dimension-five operators are troublesome, as the
Yukawa couplings must obey special relations among themselves.  One
would therefore like to avoid these operators.  Orbifold GUTs offer an
attractive solution, since they forbid dimension-five operators and,
moreover, enable us to deal with the doublet-triplet splitting problem
\cite{orbifold,altarelli01,hall01,hebecker01}.  Proton decay appears
via dimension-six operators, which can lead to unusual final states
depending on the localization of matter fields
\cite{nomura01,hebecker02}.  We will illustrate this on the basis of a
six-dimensional {\sf SO(10)} model \cite{bcw04}.  Hence, if proton
decay is observed in the future, the measured branching ratios can
make it possible to distinguish orbifold and four-dimensional GUTs.


\section{Minimal supersymmetric {\sf SU(5)}}

We start this section by briefly describing the minimal supersymmetric
{\sf SU(5)} GUT model \cite{susy-su5}.  It contains three generations
of chiral matter multiplets, \mbox{$\mathsf{10} = (Q,u^\cc,e^\cc)$},
\mbox{$\mathsf{5}^* = (d^\cc,L)$}, and a vector multiplet
$A(\mathsf{24})$ which includes the twelve gauge bosons of the SM and
twelve additional ones, the $X$ and $Y$ bosons.  Because of their
electric and color charges, the latter mediate proton decay via
dimension-six operators.  At the GUT scale, {\sf SU(5)} is broken to
${\sf G_\sm = SU(3)\times SU(2)\times U(1)_Y}$ by an adjoint Higgs
multiplet $\Sigma(\mathsf{24})$.  A pair of quintets, $H(\mathsf{5})$
and $\ol H(\mathsf{5}^*)$, then breaks ${\sf G_\sm}$ to ${\sf
  SU(3)\times U(1)_\text{em}}$ at the electroweak scale.  The
superpotential is given by
\begin{align}
  \label{minimal-superpotential}
  \begin{split}
    W & = \tfrac{1}{2} m\tr\Sigma^2 + \tfrac{1}{3} a\tr\Sigma^3 +
    \lambda \ol{H} \left(\Sigma + 3 \sigma\right) H
    + \tfrac{1}{4} Y_1^{ij} \mathsf{10}_i\, \mathsf{10}_j H + \sqrt{2}\,
    Y_2^{ij} \mathsf{10}_i\, \mathsf{5}^*_j \ol{H}
  \end{split}
\end{align}
with $m={\cal O}\left(M_\gut\right)$.  The adjoint Higgs multiplet,
\begin{align}
  \Sigma\left(\mathsf{24}\right) = \left( \begin{array}{cc}
      \Sigma_8 & \Sigma_{(3,2)} \\
      \Sigma_{(3^*,2)} & \Sigma_3
    \end{array} \right)
  + \frac{1}{2\sqrt{15}} \left( \begin{array}{cc}
      2 & 0 \\ 0 & -3
    \end{array} \right) \Sigma_{24}\ ,
\end{align}
acquires the vacuum expectation value (VEV) $\VEV{\Sigma}=\sigma \diag
\left(2,2,2,-3,-3\right)$ so that the $X$ and $Y$ bosons become
massive, whereas the SM particles remain massless.  The components
$\Sigma_8$ and $\Sigma_3$ of $\Sigma(\mathsf{24})$ both acquire the
mass
\begin{align}
  M_\Sigma\equiv M_8 = M_3 = \tfrac{5}{2} m \ .
\end{align}

The pair of quintets, $H(\mathsf{5})$ and $\ol{H}(\mathsf{5}^*)$,
contains the SM Higgs doublets, $H_f$ and $\hbf$, which break
$\mathsf{G_\sm}$, and color triplets, $H_\cc$ and $\hbcc$,
respectively.  To have massless Higgs doublets $H_f$ and $\hbf$, while
their color-triplet partners (leptoquarks) are kept super-heavy, the
mass parameters of $H(\mathsf{5})$ and $\ol{H}(\mathsf{5}^*)$ have to
be fine-tuned ${\cal O}\left(\frac{v_\ew}{\sigma}\right) \sim
10^{-13}$.  This is called the doublet-triplet-splitting problem.  The
RGE analysis gives constraints on the masses of the new particles
\cite{murayama02}.

\begin{figure}[t]
  \centering
  \subfigure[]
  { \label{fig:decay:a}
    \scalebox{.8}{
      \begin{picture}(200,60)(20,10)
        \ArrowLine(30,50)(80,30)   \Text(22,53)[]{$u_\Le$}
        \ArrowLine(30,10)(80,30)   \Text(22,12)[t]{$d_\Le$}
        \Vertex(80,30)2            \Text(80,21)[t]{$LLLL$}
        \DashLine(80,30)(130,50)5  \Text(115,60)[t]{$\widetilde l$}
        \DashLine(80,30)(130,10)5  \Text(115,12)[t]{$\widetilde q$}
        \ArrowLine(160,50)(130,50) \Text(172,52)[t]{$\nu_\Le$}
        \Vertex(130,50)1
        \Line(130,10)(130,50)
        \Photon(130,10)(130,50){2}{6}
        \Text(142,30)[]{$\widetilde{w}^\pm$}
        \Vertex(130,10)1
        \ArrowLine(160,10)(130,10) \Text(169,12)[t]{$s_\Le$}
      \end{picture}
      }
    }
  \subfigure[]
  { \label{fig:decay:b}
    \scalebox{.8}{
      \begin{picture}(180,60)(20,10)
        \ArrowLine(30,50)(80,30)   \Text(22,53)[]{$u^\cc$}
        \ArrowLine(30,10)(80,30)   \Text(22,12)[t]{$d^\cc$}
        \Vertex(80,30)2            \Text(80,21)[t]{$RRRR$}
        \DashLine(80,30)(130,50)5  \Text(110,48)[]{$\widetilde e^\cc$}
        \DashLine(80,30)(130,10)5  \Text(110,27)[]{$\widetilde u^\cc$}
        \ArrowLine(160,50)(130,50) \Text(172,52)[t]{$\nu_\Le$}
        \Vertex(130,50)1           \Text(130,57)[]{$y$}
        \ArrowLine(130,30)(130,50)
        \ArrowLine(130,30)(130,10) \Vertex(130,30)1
        \Text(142,30)[]{$\widetilde{h}^\pm$}
        \Vertex(130,10)1           \Text(128,1)[]{$y^\prime$}
        \ArrowLine(160,10)(130,10) \Text(169,12)[t]{$s_\Le$}
      \end{picture}
      }
    }
  \caption{Proton Decay via dimension-five operators: They result from
    exchange of the leptoquarks followed by gaugino or higgsino
    dressing.}
  \label{fig:decay}
\end{figure}
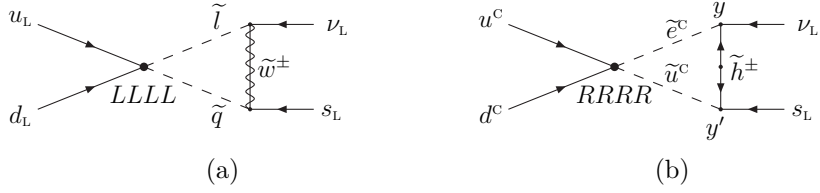

Expressed in terms of SM superfields, the Yukawa interactions are
\begin{align}
  \label{SM-superpotential}
  \begin{split}
    W_Y = \; & Y_u^{ij}\, Q_i\, u^\cc_j\, H_f + Y_d^{ij}\, Q_i\,
    d_j^\cc\, \hbf + Y_e^{ij}\, e_i^\cc\, L_j\, \hbf
    \\
    & + \tfrac{1}{2} Y_{qq}^{ij}\, Q_i\, Q_j\, H_\cc + Y_{ql}^{ij}\,
    Q_i\, L_j\, \hbcc + Y_{ue}^{ij}\, u_i^\cc\, e^\cc_j\, H_\cc +
    Y_{ud}^{ij}\, u_i^\cc\, d_j^\cc\, \hbcc \ ,
  \end{split}
\end{align}
where
\begin{align} \label{yukawa-unification}
  Y_u=Y_{qq}=Y_{ue}&=Y_1 \ , & Y_d=Y_e=Y_{ql}=Y_{ud}& =Y_2 \ .
\end{align}
Apart from the SM couplings, there are four additional ones due to the
colored leptoquarks.  Integrating out those leptoquarks, two dimension
five operators remain which lead to proton decay
(Fig.~\protect\ref{fig:decay}) \cite{dim5op},
\begin{equation}
  \label{eq:operator}
  W_5 =
  \frac{1}{M_{H_\cc}} \left[
  \, \frac{1}{2} \, Y_{qq}^{ij}\, Y_{ql}^{km}\, \left(
    Q_i\,Q_j\right)\left(
    Q_k\,L_m\right)
  + Y_{ue}^{ij}\, Y_{ud}^{km}\, \left(
    u_i^\cc\,e_j^\cc\right)\left(
    u_k^\cc\,d_m^\cc\right)\,\right] \ ,
\end{equation}
called the $LLLL$ and $RRRR$ operator.  The scalars are transformed to
their fermionic partners by exchange of a gauge or Higgs fermion.
Neglecting external momenta, the triangle diagram factor reads, up to
a coefficient $\kappa$ depending on the exchange particle,
\begin{align}
  \label{eq:exchange}
  \int \frac{d^4 k}{i (2 \pi)^4}\, \frac{1}{m_1^2 - k^2}\,
  \frac{1}{m_2^2 - k^2}\, \frac{1}{M - \SLASH[2]{k}} =
  \frac{1}{(4\pi)^2}  f(M; m_1, m_2) \ ,
  \intertext{with}
  f(M; m_1, m_2) = \frac{M}{m_1^2 - m_2^2}
  \left( \frac{m_1^2}{m_1^2 - M^2}\, \ln \frac{m_1^2}{M^2} -
  \frac{m_2^2}{m_2^2-M^2}\, \ln \frac{m_2^2}{M^2} \right)\ ,
  \label{eq:triangle}
\end{align}
where $M$ and $m_j$ denote the gaugino and sfermion masses,
respectively.

As a result of Bose statistics for superfields, the total
anti-symmetry in the colour index requires that these operators are
flavor non-diagonal \cite{dimopoulos82}.  The dominant decay mode is
therefore $p\to K \bar\nu$.  Since the dressing with gluinos and
neutralinos is flavor diagonal, the chargino exchange diagrams are
dominant \cite{nath85,hisano93}.  The wino exchange is related to the
$LLLL$ operator and the charged higgsino exchange to the $RRRR$
operator, so that the coefficients of the triangle diagram factor are
\begin{align}
  \label{eq:kfactor}
  \kappa_\Le & = 2g^2 \ , & \kappa_\ri & = y\, y^\prime \ .
\end{align}
Here $y$ and $y^\prime$ denote the corresponding Yukawa couplings (cf.
Fig.~\ref{fig:decay:b}) and $g$ is the gauge coupling.

Let us sketch how to calculate the decay rate via dimension-five
operators; for details see Ref.~\cite{cw03}.  The Wilson coefficients
$C_{5\Le} = Y_{qq} Y_{ql}$ and $C_{5\ri} = Y_{ue} Y_{ud}$ are
evaluated at the GUT scale.  Then they have to be evolved down to the
SUSY breaking scale, leading to a short-distance renormalization
factor $A_s$.  Now the sparticles are integrated out, and the
operators give rise to the effective four-fermion operators of
dimension 6.  The renormalization group procedure goes on to the scale
of the proton mass, $m_p\sim 1\,$GeV, leading to a second,
long-distance renormalization factor $A_l$ \cite{ellis82}.  At 1\,GeV,
the link to the hadronic level is made using the chiral Lagrangean
method \cite{chiral}.  Thus the decay width can be written as
\begin{align}
  \Gamma & = \sum \left| \, {\cal K}_\text{had}\, A_l\,
    \frac{\kappa}{(4\pi)^2} f(M; m_1, m_2)\, A_s\,
    \frac{1}{M_{H_\cc}}\, C_5 \right|^2 \ ,
\end{align}
where the sum includes all possible diagrams.

The decay width for the dominant channel $p\to K^+ \bar\nu$ reads
\begin{multline} \label{width-kaon}
  \Gamma (p\to K^+ \bar\nu) = \dfrac{(m_p^2-m_K^2)^2}{32\pi m_p^3
    f_\pi^2} \sum_\nu \left| \, C_5^{usd\nu}\, \frac{2 m_p}{3 m_B} D +
    \, C_5^{uds\nu}\, \left( 1+\frac{m_p}{3 m_B}(3F+D) \right)
  \right. \\
  + \left. C_5^{dsu\nu}\, \left( 1-\frac{m_p}{3 m_B}(3F-D) \right)
  \right|^2 \ .
\end{multline}
Here, $m_p$ and $m_K$ denote the masses of the proton and kaon,
respectively, and $f_\pi$ is the pion decay constant.  $m_B$ is an
average baryon mass according to contributions from diagrams with
virtual $\Sigma$ and $\Lambda$ \cite{chiral}.  $D$ and $F$ are the
symmetric and antisymmetric {\sf SU(3)} reduced matrix elements for
the axial-vector current.

According to the two Wilson coefficients, the coefficients $C_5$ split
into two parts,
\begin{align}
  C_5 = \beta\, C_{\Le\Le} + \alpha\, C_{\ri\Le} \ ,
\end{align}
with
\begin{align}
  \label{eq:coefficient}
  C_{\Le\Le/\ri\Le} = \frac{1}{M_{H_\cc}}\, C_{5\Le/5\ri}\, A_s \,
  A_l\, \frac{\kappa_{\Le/\ri}}{(4\pi)^2} f(M; m_1, m_2)
\end{align}
and the hadron matrix elements $\alpha$ and $\beta$ \cite{brodsky83},
\begin{align} \label{eq:alphabeta}
  \alpha\, P_\Le\, u_p & = \eabg \braket{0}{\left( d_\ri^\alpha\,
      u_\ri^\beta \right) u_\Le^\gamma}{\,p} \ , & \beta\, P_\Le\, u_p
  & = \eabg \braket{0}{\left( d_\Le^\alpha\, u_\Le^\beta \right)
    u_\Le^\gamma}{\,p} \ .
\end{align}

While the hadronic parameters are fairly known, the masses and mixings
of the SUSY-particles are unknown.  We know through their absence that
they have to be heavier than ${\cal O}\,(100\,\text{GeV})$; on the
other hand, they are expected not to be much heavier than ${\cal
  O}\,(1\,\text{TeV})$. Looking at the dressing diagram we notice that
when taking the sfermions to be degenerate at a TeV, the triangle
diagram factor (\ref{eq:triangle}) is given by
\begin{equation}
  \label{triangle2}
  f(M;m) =
  \dfrac{M}{(M^2-m^2)^2} \left(m^2-M^2-M^2\, \ln\frac{m^2}{M^2}\right)
  \; \xrightarrow{\; M \ll m\;} \;
  \frac{M}{m^2} \ .
\end{equation}
The simplest case is to assume that the sfermions have masses of
1\,TeV.  An exception is often made for top squarks.  Since the
off-diagonal entries of the mass matrix are proportional to $m_t$, the
mixing in the stop sector is expected to be large, with at least one
eigenvalue much below 1\,TeV.  In analyses, one typically uses
400\,GeV, 800\,GeV, or 1\,TeV for $m_{\tilde t}$.  For the other
sfermions, the mixings are neglected.  The proton decay rate is
further suppressed by light gauginos and higgsinos.  Note that the
experimental limit for charginos is
$m_{\widetilde{\chi}^\pm}>67.7\,$GeV \cite{pdg}.

Since proton decay is dangerously large, the decoupling scenario
\cite{decoupl} has also been studied, where the scalars of the first
and second generation can be as heavy as 10\,TeV \cite{murayama02}.
Such an adjustment has been motivated by the supersymmetric flavor
problem.  In this scenario, proton decay via dimension-five operators
is clearly dominated by the third generation.

\medskip

Let us analyze the Wilson coefficients now.  Minimal {\sf SU(5)}
predicts that the Yukawa couplings of down quarks and charged leptons
are unified, as can be seen from Eqs.~(\ref{yukawa-unification}).
While $m_b=m_\tau$ can be fulfilled at the GUT scale, the equalities
fail for the first and second generation.  Nevertheless, proton decay
via dimension-five operators has been analyzed assuming
\begin{align} \label{eq:minimal-assumption}
  Y_{qq}=Y_{ue}&=Y_u\ , &  Y_{ql}=Y_{ud}&=Y_d \ .
\end{align}
Then the decay rate can be calculated as described above.  Note,
however, that the choices $Y_{ql}=Y_{ud}=Y_e$ or $Y_{ql}=Y_d$,
$Y_{ud}=Y_e$ would be equally justified.  As we shall see, this
ambiguity strongly affects the proton decay rate.\footnote{Another
  uncertainty concerns the sfermion mixings.  Due to constraints by
  flavor changing neutral currents, they are assumed to coincide with
  the fermion mixings; see, however Refs.~\cite{bajc02}.}

\begin{figure}[t]
  \centering \includegraphics[width=.68\linewidth]{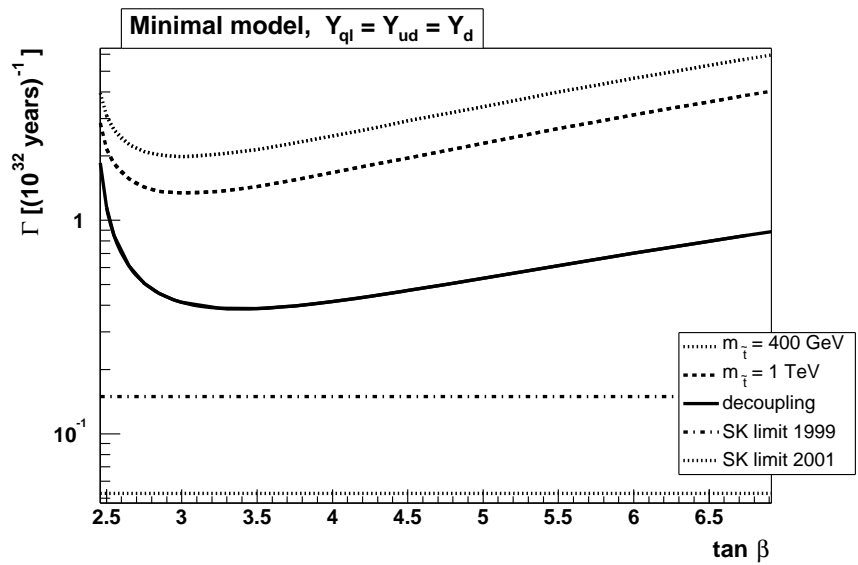}
  \caption{Decay rate $\Gamma(p\to K^+ \bar\nu)$ as function of
    $\tan\beta$ in minimal {\sf SU(5)} with
    \mbox{$Y_{ql}=Y_{ud}=Y_d$}.}
  \label{fig:minimal-down}
\end{figure}

Fig.~\ref{fig:minimal-down} shows the results of the following three
cases: (i) all sfermions have masses of 1\,TeV; (ii) $m_{\tilde t}$ is
changed to 400\,GeV; (iii) decoupling scenario, where the scalars of
the first and second generation have masses of 10\,TeV.  The
dash-dotted line represents the experimental limit $\tau = 6.7\times
10^{32}$\,years as given by the Super-Kamiokande experiment
\cite{pdg,hayato99}, the dotted line is the newer limit $\tau =
1.9\times 10^{33}$\,years \cite{ganezer01}.

The Wilson coefficients are proportional to \mbox{$\tan\beta +
  \frac{1}{\tan\beta}$} and \mbox{$(\tan\beta +
  \frac{1}{\tan\beta})^2$}, where $\tan\beta$ is the ratio of the
vacuum expectation values of the Higgs doublets $H_f$ and $\hbf$.
Clearly low values are preferred for obtaining a low decay rate; on
the other hand, the top Yukawa coupling becomes non-perturbative for
$\tan\beta\lesssim 2.5$.

As already mentioned, it is possible to constrain the leptoquark mass
using the RGEs.  These constraints depend strongly on the Higgs
representations, therefore we choose the most conservative value
$M_{H_\cc}=M_\text{GUT}=2\times 10^{16}\,\text{GeV}$ in order to study
whether the experimental limit already rules out this model.

The decay rate in Fig.~\ref{fig:minimal-down} is always above the
experimental limit, which led to the claim that minimal {\sf SU(5)} is
excluded \cite{goto99,murayama02}.  But as already discussed, there is
no compelling reason for the assumption $Y_{ql}=Y_{ud}=Y_d$!  In order
to illustrate the strong dependence of the decay rate on flavor mixing
and therefore on Yukawa unification, let us study the case
\begin{align}
  Y_{qq}=Y_{ue}&=Y_u \ , & Y_{ql}=Y_{ud}&=Y_e \ .
\end{align}
Then the mixing matrix $U_u^\dagger\,U_d\equiv V_\ckm$, which appears
in the Wilson coefficients, is replaced by ${\cal
  M}=U_u^\dagger\,U_e$.  Note that the mixing matrix in $Y_u$ or $Y_d$
is still given by $V_\ckm$.  Since $Y_d\not= Y_e$, the masses and
mixing of quarks and leptons are different and ${\cal M}$ is
undetermined.

\begin{figure}
  \centering
  \includegraphics[width=.68\linewidth]{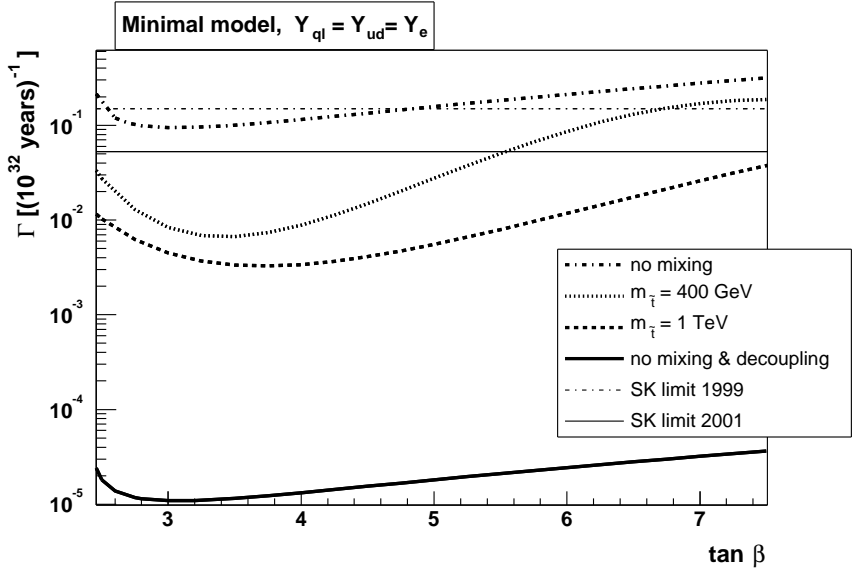}
  \caption{Decay rate $\Gamma(p\to K^+ \bar\nu)$ as a function of
    $\tan\beta$ with $Y_{ql}=Y_{ud}=Y_e$.}
  \label{fig:minimal-electron}
\end{figure}

We first ignore mixing, i.\,e. ${\cal M}=\mathbbm{1}$, and calculate
the decay rate; the results are shown in
Fig.~\ref{fig:minimal-electron}.  Without mixing, only scalars of the
first and second generation take part, so the decay rate is
significantly reduced in the decoupling scenario where the triangle
diagram factor (\ref{eq:triangle}) changes by almost two orders of
magnitude.

If, moreover, we take ${\cal M}$ arbitrarily, it is possible to push
the decay width below the experimental limit even for smaller sfermion
masses.  In the case $m_{\tilde t}=400$\,GeV, this is only possible
for small values of $\tan\beta$.

The fact that a sufficiently low decay rate can be found illustrates
the dependence on flavor mixing and therefore the uncertainty due to
the failure of Yukawa unification.  Minimal supersymmtric {\sf SU(5)}
can only be excluded by the mismatch between the Yukawa couplings of
down quarks and charged leptons as non-supersymmetric {\sf SU(5)} is
excluded by the failure of gauge unification.


\section{Consistent supersymmetric {\sf SU(5)}}

Minimal supersymmetric {\sf SU(5)} is inconsistent due to the failure
of Yukawa unification, thus additional interactions are required which
account for the difference of down quark and charged lepton masses.
Such interactions are conveniently parameterized by higher dimensional
operators, which are naturally expected as a result of interactions at
a higher scale, where the GUT model is extended to a more fundamental
theory.  In particular, the GUT scale is only about two orders below
the Planck scale.

Including possible terms up to order $1/M_\text{Pl}$, the
superpotential reads \cite{bajc02b}
\begin{align} \label{eq:consistent-w}
  W & = \tfrac{1}{2}m\tr\Sigma^2 + \tfrac{1}{3}a\tr\Sigma^3 + b
  \frac{(\tr\Sigma^2)^2}{M_\text{Pl}} + c
  \frac{\tr\Sigma^4}{M_\text{Pl}}
  \nonumber\\[1pt]
  & \quad + \frac{1}{4}\,\epsilon_{abcde}\left( Y_1^{ij}
    \mathsf{10}_i^{ab} \mathsf{10}_j^{cd} H^e + f_1^{ij}
    \mathsf{10}_i^{ab} \mathsf{10}_j^{cd}
    \frac{\Sigma^e_f}{M_\text{Pl}} H^f + f_2^{ij} \mathsf{10}_i^{ab}
    \mathsf{10}_j^{cf}\, H^d \frac{\Sigma^e_f}{M_\text{Pl}} \right)
  \nonumber \\
  & \quad + \sqrt{2}\left( Y_2^{ij} \ol{H}_a \mathsf{10}_i^{ab}
    \mathsf{5}^*_{jb} + h_1^{ij} \ol{H}_a
    \frac{\Sigma^a_b}{M_\text{Pl}} \mathsf{10}_i^{bc}
    \mathsf{5}^*_{jc} + h_2^{ij} \ol{H}_a \mathsf{10}_i^{ab}
    \frac{\Sigma_b^c}{M_\text{Pl}} \mathsf{5}^*_{jc} \right) \ .
\end{align}
Now the masses of $\Sigma_3$ and $\Sigma_8$ are no longer identical,
affecting the constraint on the leptoquark mass
\cite{bajc02b}.

The Yukawa couplings are then given by
\begin{align}
  \label{sm-planck}
  \begin{split}
    Y_u & = Y_1 + 3\frac{\sigma}{M_\text{Pl}}f_1^S +
    \frac{1}{4}\frac{\sigma}{M_\text{Pl}}\left(3f_2^S+5f_2^A\right)
    \ , \\
    Y_d & = Y_2 - 3\frac{\sigma}{M_\text{Pl}}h_1 +
    2\frac{\sigma}{M_\text{Pl}}h_2
    \ , \\
    Y_e & = Y_2 -3\frac{\sigma}{M_\text{Pl}}h_1 -
    3\frac{\sigma}{M_\text{Pl}}h_2 \ .
  \end{split}
\end{align}
Here $\sigma/M_\text{Pl}\sim{\cal O}\left(10^{-2}\right)$, and $S$ and
$A$ denote the symmetric and antisymmetric parts of the matrices.
Thus the three Yukawa matrices, which are related to masses and mixing
angles at $M_Z$ by the RGEs, are determined by six matrices.  From
Eqn.~(\ref{sm-planck}) one reads off
\begin{align}
  \label{c-relation}
  Y_d - Y_e = 5\frac{\sigma}{M_\text{Pl}}h_2 \ ,
\end{align}
hence, the failure of Yukawa unification is naturally accounted for by
the presence of $h_2$.  Note that no additional field is introduced to
obtain this relation; it just arises from corrections ${\cal
  O}\left(\sigma/M_\text{Pl}\right)$.

\medskip

Let us study the effect of those higher-dimensional operators on the
Wilson coefficients.  It is instructive to express the couplings in
terms of the quark and charged lepton Yukawa couplings and the
additional matrices $f$ and $h$,
\begin{align}
  \label{eq:wilson-planck-f}
    Y_{qq} = Y_{qq}^S = Y_{ue}^S &= Y_u^S -
    5\frac{\sigma}{M_\text{Pl}}\left( f_1^S+\frac{1}{4} f_2^S\right) ,
    & Y_{ql} & = Y_e + 5\frac{\sigma}{M_\text{Pl}}h_1 \ ,
  \\
  \label{eq:wilson-planck-h}
  Y_{ue}^A & = Y_u^A - \frac{5}{2}\,\frac{\sigma}{M_\text{Pl}}f_2^A \
  , & Y_{ud} & = Y_d + 5\frac{\sigma}{M_\text{Pl}}h_1 \ .
\end{align}
Note that $Y_{ql}-Y_{ud}=Y_e-Y_d$, which means that $Y_{ql}$ and
$Y_{ud}$ cannot be zero at the same time.  Nevertheless, proton decay
via dimension-five operators can be avoided if both
$C_{5L}=Y_{qq}Y_{ql}$ and $C_{5R}=Y_{ue}Y_{ud}$ vanish.  For this
purpose the couplings have to fulfill the relations
\begin{align}
  \label{eq:d5-avoid}
    f_1^S + \frac{1}{4} f_2^S
    & = \frac{M_\text{Pl}}{5\,\sigma}\, Y_u^S \ ,
    & f_2^A & = \frac{2}{5}\,\frac{M_\text{Pl}}{\sigma}\, Y_u^A \ .
\end{align}
This is only possible if we allow the (3,3)-component of $f_1$ and
$f_2$ to be ${\cal O} \left(\frac{M_\text{Pl}}{\sigma}\right)\gg 1$.
But even if we restrict ourselves to `natural matrices', i.\,e.
couplings up to ${\cal O} \left( 1 \right)$, we can considerably
reduce the decay amplitudes.  We will illustrate this with two simple
examples where either the $RRRR$ or the $LLLL$ contribution vanishes
at the GUT scale.

\begin{figure}[t]
  \centering
  \includegraphics[width=.8\linewidth]{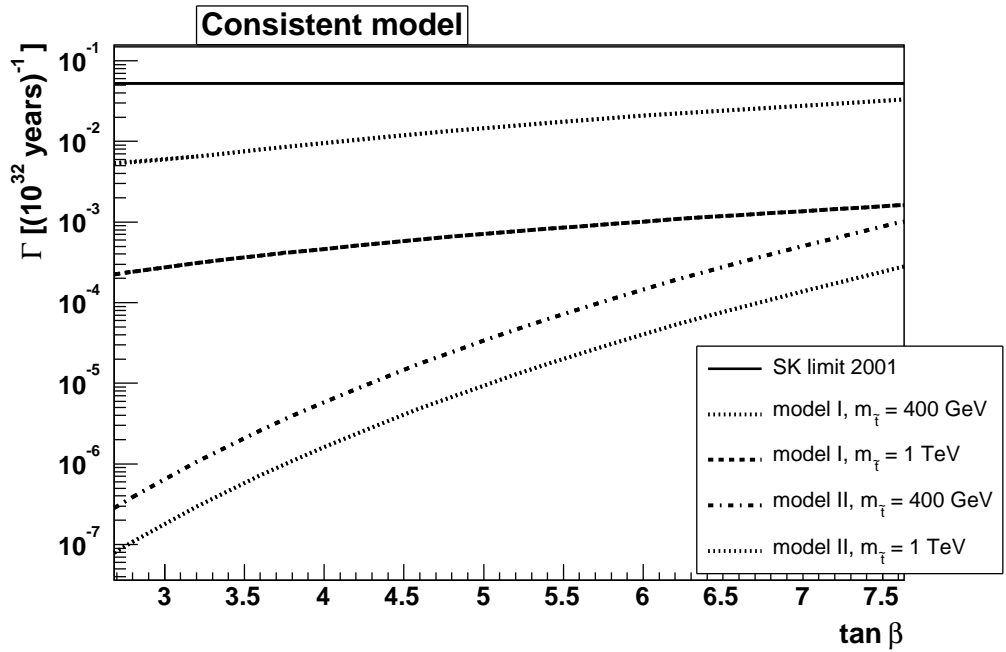}
  \caption{Decay rate $\Gamma(p\to K^+ \bar\nu)$ as function of
    $\tan\beta$ in the consistent models.
    \label{fig:consistent}}
\end{figure}

We assume that $Y_{qq}$, $Y_{ql}$, $Y_{ue}$ and $Y_{ud}$ are all
diagonal by a suitable choice of matrices.  The simplest form of
$Y_{qq}$ and $Y_{ue}$ is then
\begin{align}
  \label{eq:consistent-model}
    Y_{qq} = Y_{ue} & = \diag \left(0,0,y_t \right) \ ,
\end{align}
where $y_t$ is the top Yukawa coupling at $M_\gut$.
In the first model, we spread $Y_e-Y_d$ such that
\begin{align}
  \label{eq:consistent-model1}
  Y_{ud} & = \diag \left(0,y_s-y_\mu,y_b-y_\tau \right) \ , & Y_{ql} &
  = \diag \left(y_e-y_d,0,0 \right) \ .
\end{align}
Clearly $C_{5R}^{ijkm}=Y_{ue}^{ij}\ Y_{ud}^{km}$ is zero whenever a
particle of the first generation takes part.  Since at least one
particle of the first generation is needed, the $RRRR$ contribution
vanishes completely.  After RGE evolution, this simple structure of
Wilson coefficients changes slightly, but as shown in
Fig.~\ref{fig:consistent}, the decay amplitude is always well below
the experimental limit --- in the case $m_{\tilde t}=1$\,TeV even more
than two orders of magnitude.  Here, we restrict ourselves to the
cases where the sfermion masses are not heavier than 1\,TeV.

If we choose the matrices $Y_{ql}$ and $Y_{ud}$ as
\begin{align}
  Y_{ud} &= \diag \left(y_d-y_e,y_s-y_\mu,y_b \right)\ , & Y_{ql} &=
  \diag \left(0,0,y_\tau \right)\ ,
\end{align}
the $LLLL$ contribution vanishes at $M_\gut$ because now
$C_{5L}^{ijkm}=Y_{qq}^{ij}\ Y_{ql}^{km}$ is only different from zero
for $i=j=k=m=3$, but the decay has to be non-diagonal.  Only the
$RRRR$ contribution with a low absolute value remains.  After
renormalization, this contribution is still dominated by third
generation scalars, and the $LLLL$ operator contributes only via $p\to
K\bar\nu_{\tau}$. In this case, the decay rate is even smaller
(Fig.~\ref{fig:consistent}).  Furthermore, due to the smaller
(3,3)-component of $h_1$ compared to the first model, it can easily be
used for higher values of $\tan\beta$.

\medskip

We have seen that higher-dimensional operators can reduce the proton
decay rate by several orders of magnitude and make it consistent with
the experimental upper bound.  This impressing fact leads to the
question, whether there is any mechanism which would naturally lead to
the required relations among Yukawa couplings.  We can think of two
possibilities, to start with some ad-hoc textures as a result of an
unknown additional symmetry \cite{textures} or to extend the analysis
to another group, in order to obtain additional symmetry restrictions.
{\sf SO(10)}, however, does not restrict the contributions from the
higher-dimensional operators \cite{bcw04b}.  Thus it is more promising
to consider theories, where the dimension-five operators are
generically absent.


\section{Orbifold GUTs}

In orbifold GUTs, the GUT gauge symmetry is realized in more than four
space-time dimensions and broken to the standard model by
compactification on an orbifold, utilizing boundary conditions that
violate the GUT symmetry
\cite{orbifold,altarelli01,hall01,hebecker01}.  As a consequence, the
GUT and electroweak scale are separated in an elegant way and the
dimension-five operators are absent \cite{altarelli01,hall01}.
Furthermore, in case of {\sf SO(10)}, the breaking to $G_\sm$ via the
Higgs mechanism requires large Higgs representations and the path is
not unique.  Here, orbifold GUTs provide an attractive solution as
well, as orbifold symmetry breaking can simplify the breaking pattern.

\begin{figure}[t]
  \centering \scalebox{.45}{
    \begin{picture}(200,200)(0,0)
      \Curve{(10,10)(100,25)(190,10)}
      \Curve{(10,190)(100,175)(190,190)}
      \begin{sideways}
        \Curve{(10,-10)(100,-25)(190,-10)}
        \Curve{(10,-190)(100,-175)(190,-190)}
      \end{sideways}
      \Vertex(10,10)4
      \Text(0,10)[r]{\huge $O \left[{\sf SO(10)}\right]$}
      \Vertex(10,190)4
      \Text(0,190)[r]{\huge $O_\GG \left[{\sf G_\GG}\right]$}
      \Vertex(190,190)4
      \Text(200,190)[l]{\huge $O_\text{fl} \left[{\sf
            G_\text{fl}}\right]$}
      \Vertex(190,10)4
      \Text(200,10)[l]{\huge $O_\ps \left[{\sf G_\ps}\right]$}
    \end{picture}
    }
  \caption{The three {\sf SO(10)} subgroups at the corresponding
    fixed points (branes) of the orbifold \mbox{$T^2/{\mathbbm Z}_2
      \times {\mathbbm Z}_2^\prime \times {\mathbbm
        Z}_2^{\prime\prime}$}.
    \label{fig:orb}}
  \vspace{-10pt}
\end{figure}
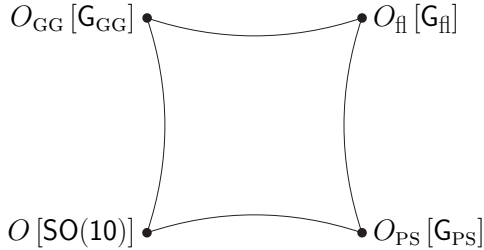
We consider a 6D SUSY {\sf SO(10)} model compactified on a torus with
three ${\mathbbm Z}_2$ parities \cite{so10-6d}.  The first ${\mathbbm
  Z}_2$ breaks only $N=1$ 6D SUSY to $N=1$ 4D SUSY, the remaining ones
{\sf SO(10)} as well, namely to ${\sf G_\GG=SU(5)\times U(1)}$ and ${\sf
  G_\ps = SU(4)\times SU(2)\times SU(2)}$ \cite{PS}, respectively.
Then the zero modes belong to the intersection of the two symmetric
subgroups, ${\sf G_\sm\times U(1)^\prime}$.  The physical region is a
`pillow', corresponding to the two compact dimensions, with four fixed
points as corners, where the unbroken subgroups are {\sf SO(10)},
${\sf G_\ps}$, ${\sf G_\GG}$, and flipped {\sf SU(5)} \cite{flipped}
(Fig.~\ref{fig:orb}).

The field content is as follows \cite{asaka03}: The matter fields are
located on the branes, whereas the gauge fields, six 10-dimensional
Higgs representations and two combinations ${\sf 16\oplus \ol{\sf
    16}}$, are bulk fields.  The matter fields include the {\sf 10}
and ${\sf 5}^*$ of {\sf SU(5)} plus a singlet, the right-handed
neutrino.  With this setup, both the irreducible and reducible 6D
gauge anomalies vanish.

The main idea to generate fermion mass matrices is to locate the three
sequential {\sf 16}-plets on the three branes where {\sf SO(10)} is
broken to its three GUT subgroups, namely $\psi_1$ at $O_\GG$,
$\psi_2$ at $O_\text{fl}$ and $\psi_3$ at $O_\ps$.  The three
`families' are then separated by distances large compared to the
cutoff scale $M_*={\cal O}\left(10^{17}\,\text{GeV}\right)$, where the
theory becomes strongly coupled.  Thus they can only have diagonal
Yukawa couplings with the bulk Higgs fields, direct mixings are
exponentially suppressed.  The brane fields mix with the bulk field
zero modes.  These mixings take place only among left-handed leptons
and right-handed down-quarks, which leads to a characteristic pattern
of mass matrices.

The mass terms read
\begin{align}
  W = d_\alpha m^d_{\alpha\beta} d^\cc_\beta + e^\cc_\alpha
  m^e_{\alpha\beta} e_\beta + n^\cc_\alpha m^D_{\alpha\beta} \nu_\beta
  + u_i m^u_{ij} u^\cc_j + \tfrac{1}{2} n^\cc_i m^N_{ij} n^\cc_j \ .
\end{align}
Here $m^u$ and $m^N$ are diagonal $3\times 3$ matrices but due to the
mixing with the bulk field zero modes, $m^d$, $m^e$ and $m^D$ are
$4\times 4$ matrices with a lopsided structure,
\begin{align} \label{lopsided-matrix}
  \frac{1}{\tan{\beta}}\, m^u \sim \frac{v_u M_*}{v_{B-L}^2}\, m^N &
  \sim \diag\left( \mu_1,\mu_2,\mu_3 \right) ,
  \\
  \frac{1}{\tan{\beta}}\, m^D \sim m^d \sim m^e & \sim \left(
    \begin{array}{cccc}
      \mu_1 & 0 & 0 & \tilde \mu_1 \\
      0 & \mu_2 & 0 & \tilde \mu_2 \\
      0 & 0 & \mu_3 & \tilde \mu_3 \\
      \widetilde M_1 & \widetilde M_2 & \widetilde M_3 &
      \widetilde M_4
    \end{array}
  \right) ;
\end{align}
$\mu_i$ and $\tilde\mu_j$ are ${\cal O}\,(v_\ew)$, whereas $\widetilde
M_\alpha$ are of the order of the compactification scale.

The fermion masses and mixings agree well with the data within
coefficients ${\cal O}\left(1\right)$ in the case $\mu_{1,2}\to 0$
\cite{asaka03}.  Note that while $m_\tau \simeq m_b$, the muon and
electron masses can be easily accommodated because $\mu_2^d$ and
$\mu_2^e$ are not related by a flipped {\sf SU(5)} mass relation.

While the mixings for the left-handed down quarks and right-handed
charged leptons are small so that their diagonalization matrices read
\begin{align}
  U^d_\Le = V_\ckm \simeq U^e_\ri \ ,
\end{align}
those for the right-handed down quarks and left-handed charged leptons
are large.  For those, the diagonalization matrices are given by
\begin{align} \label{eq:down-right}
  U^d_\ri \sim U^e_\Le \simeq \left( \begin{array}{cccc}
      -\frac{\widetilde{M}_2}{\widetilde{M}_{12}} &
      \frac{\widetilde{M}_1 \left( \tilde\mu_3
          \widetilde{M}_3-\mu_3\widetilde{M}_4 \right)}{\bar\mu\,
        \widetilde{M}\, \widetilde{M}_{12}} & -\frac{\widetilde{M}_1
        \left( \tilde\mu_3 \widetilde{M}_4+\mu_3 \widetilde{M}_3
        \right)}{\bar\mu\, \widetilde{M}^2} &
      \frac{\widetilde{M}_1}{\widetilde{M}}
      \\
      \rule[-4mm]{0mm}{10mm}%
      \frac{\widetilde{M}_1}{\widetilde{M}_{12}} &
      \frac{\widetilde{M}_2 \left( \tilde\mu_3 \widetilde{M}_3-\mu_3
          \widetilde{M}_4 \right)}{\bar\mu\, \widetilde{M}\,
        \widetilde{M}_{12}} & -\frac{\widetilde{M}_2 \left(
          \tilde\mu_3 \widetilde{M}_4+\mu_3 \widetilde{M}_3
        \right)}{\bar\mu\, \widetilde{M}^2} &
      \frac{\widetilde{M}_2}{\widetilde{M}}
      \\
      \rule[-5mm]{0mm}{10mm}%
      0 & -\frac{\tilde\mu_3}{\bar\mu}
      \frac{\widetilde{M}_{12}}{\widetilde{M}} & -\frac{\tilde\mu_3
        \widetilde{M}_3\widetilde{M}_4-\mu_3 \left(
          \widetilde{M}_1^2+\widetilde{M}_2^2+\widetilde{M}_4^2
        \right)}{\bar\mu\, \widetilde{M}^2} &
      \frac{\widetilde{M}_3}{\widetilde{M}}
      \\
      0 & \frac{\mu_3}{\bar\mu}
      \frac{\widetilde{M}_{12}}{\widetilde{M}} & \frac{\tilde\mu_3
        \left( \widetilde{M}_1^2+\widetilde{M}_2^2+\widetilde{M}_3^2
        \right)-\mu_3 \widetilde{M}_3\widetilde{M}_4}{\bar\mu\,
        \widetilde{M}^2} & \frac{\tilde M_4}{\widetilde{M}}
    \end{array} \right)
\end{align}
up to a two-dimensional mixing matrix for the second and third
generation, which can be neglected for the purpose of proton decay.
In Eqn. (\ref{eq:down-right}), $\bar\mu_3$ is ${\cal O} \left( \mu_3,
  \tilde\mu_3 \right)$, furthermore
$\widetilde{M}_{12}=\sqrt{\widetilde{M}_1^2+\widetilde{M}_2^2}$ and
$\widetilde M^2 = \sum_\alpha \widetilde M_\alpha^2$.

\medskip

The up-quarks are confined to one fixed point each, in particular the
up quark is located on the Georgi-Glashow one.  Therefore
dimension-six proton decay can arise via the exchange of the {\sf
  SU(5)} $X$ and $Y$ bosons as in the four-dimensional case,
\begin{align} \label{effWeyl-RRLL}
  {\mathscr L}_\text{eff} &= - \frac{g_5^2 }{ M_V^2}\;
  \epsilon_{\alpha\beta\gamma} \left[ \ol{e^c}_{\!j}
    \ol{u^c}_{\!\alpha i}\, Q_{\beta i}\, Q_{\gamma,j} -
    \ol{d^c}_{\!\alpha k} \ol{u^c}_{\!\beta i}\, Q_{\gamma i}\, L_k
  \right] + \text{h.c.}  \, .
\end{align}

The analysis of these operators is analogous to the dimension-five
case; for details see Ref.~\cite{bcw04}.  Contrary to the 4D case, we
have to take into account the presence of the Kaluza-Klein tower of
the vector bosons $V=\left(X,Y\right)$ with masses
\begin{align}
  M_V^2 (n,m) = \frac{(2n+1)^2}{R_5^2} + \frac{(2m)^2}{R_6^2} \ .
\end{align}
Furthermore, the KK modes interact more strongly by a factor of
$\sqrt{2}$ due to their bulk normalization \cite{hall01}.

The sum over the KK modes is logarithmically divergent.  Since the
theory is only valid below the scale $M_*$, we restrict the sum to
masses $M_V (n,m)\leq M_*$. In the case $R_5 = R_6 = 1/M_c$, we obtain
\begin{align}  \label{eq:sum-kin}
  \frac{1}{(M_V^\text{eff})^2} = 2 \sum_{n,m=0}^\infty \frac{1}{M_V^2
    (n,m)} \simeq \frac{\pi}{4\,M_c^2}\, \left( \ln \left(
      \frac{M_*}{M_c} \right) + 2.3 \right) \ .
\end{align}
As we will see below, $M_c$ is constrained by the experimental limit
of the dominant decay mode $p\to e^+ \pi^0$ yielding $M_*/M_c\simeq
12$.

Due to the parities and the ${\sf SO(10)}$ breaking pattern, the
coupling of the gauge bosons is not universal any more, in contrast to
the 4D case.  Their wavefunctions vanish on the fixed points $O_\ps$
and $O_\text{fl}$ so that no coupling arises via the kinetic term with
the second and third generation.  The couplings to the bulk states are
irrelevant since these are embedded in full {\sf SU(5)} multiplets
together with massive KK modes, thus the interaction always mixes the
light states with the heavy ones \cite{nomura01,hebecker02}.

Altogether, the operators for the decay via $X$ and $Y$ bosons read
\begin{multline}  \label{eq:eff6D-VU}
  {\mathscr L}_\text{eff} = \frac{g_5^2}{(M_V^\text{eff})^2} \left[\,
    \ol{d^\cc}_{\!\!\alpha,l} \left( U^{d\top}_\ri \right)_{l1}
    \ol{u^\cc}_{\!\beta 1} \left( u_{\gamma 1} \left( U^e_\Le
      \right)_{1j} e_j - d_{\gamma m} \left( U^d_\Le \right)_{1m}
      \left(U^{\nu}_\Le \right)_{1j} \nu_j \right) \right.
  \\
  + \left. 2\, \ol{e^\cc}_{\!k} \left( U^{e\top}_\ri \right)_{k1}
    \ol{u^\cc}_{\!\!\alpha 1} \, d_{\beta m} \left( U^d_\Le
    \right)_{1m} u_{\gamma 1} \vphantom{\left( u_{\gamma 1} \left(
          U^e_\Le \right)_{1j} \right)} \right] \eabg + \text{h.c.}
\end{multline}
with the fermions in their mass eigenstates.

We start with the simplest case of $U_\ri^d=U_\Le^e$ and degenerate
masses $\widetilde M_\alpha$ in the limit $\tilde\mu_3=\mu_3$, which
we denote case I.  In this case, the state $d_{R1}$ has no
strange-component and we obtain
\begin{multline}
  {\mathscr L}_\text{eff} = \frac{g_5^2}{(M_V^\text{eff})^2}\; \eabg
  \left[ 2\,V_{ud}^2\, \ol{e^\cc}\, \ol{u^\cc_\alpha}\, d_\beta\,
    u_\gamma +\frac{1}{2}\, \ol{d^\cc_\alpha}\, \ol{u^\cc_\beta}\,
    u_\gamma\, e + 2\,V_{ud} V_{us}\, \ol{\mu^\cc}\,
    \ol{u^\cc_\alpha}\, d_\beta\, u_\gamma \vphantom{\sum_{j=1}^3}
  \right.
  \\
  + 2\,V_{ud} V_{us}\, \ol{e^\cc}\, \ol{u^\cc_\alpha}\, s_\beta\,
  u_\gamma + 2\,V_{us}^2\, \ol{\mu^\cc}\, \ol{u^\cc_\alpha}\,
  s_\beta\, u_\gamma
  \\[-4pt]
  - \left. \sum_{j=1}^3 \frac{1}{\sqrt{2}}\, \left( U^\nu_L
    \right)_{1j}\, \ol{u^\cc_\alpha}\, \ol{d^\cc_\beta}\, \left\{
      V_{ud}\, d_\gamma\, + V_{us}\, s_\gamma
      \vphantom{\frac12}\right\}\, \nu_j \right] + \text{h.c.}
\end{multline}
The numerical results for the branching ratios are shown in
Table~\ref{tb:result}.

Additional operators can arise at any brane from couplings which
contain the derivative along the extra dimensions of the locally
vanishing gauge bosons \cite{hebecker02}.  Their contribution is
suppressed with respect to the standard operators \cite{bcw04}; the
coefficients, however, are undetermined.  For $c_5=c_6=1$, the
corrections for the listed branching ratios are less than 3\%.

\begin{table}[t]
  \centering
  \begin{tabular}{l|rr|r}
    decay channel & \multicolumn{3}{c}{Branching Ratios [\%]} \\
    \cline{2-4}
    & \multicolumn{2}{c|}{6D {\sf SO(10)}} &
    4D {\sf SU(5)} \\
    & case I & case II & \\
    \hline
    $e^+\pi^0$ & 75 & 71 & 54 \\
    $\mu^+\pi^0$ & 4 & 5 & \textless\,1 \\
    $\bar\nu\,\pi^+$ & 19 & 23 & 27 \\
    $e^+ K^0$ & 1 & 1 & \textless\,1 \\
    $\mu^+ K^0$ & \textless\,1 & \textless\,1 & 18 \\
    $\bar\nu\, K^+$ & \textless\,1 & \textless\,1 & \textless\,1 \\
    \hline
  \end{tabular}
  \caption{Resulting branching ratios and comparision with
    \mbox{${\sf SU(5)}\times {\sf U(1)}_F$}. \label{tb:result}}
\end{table}

To make a comparison with ordinary 4D GUT models, we consider two {\sf
  SU(5)} models described in Refs.~\cite{sato99,altarelli04}, where we
assume that some mechanism suppresses or avoids the proton decay
arising from dimension-five operators.\footnote{For recent discussions
  of dimension-six operators in flipped {\sf SU(5)} and {\sf SO(10)},
  see Refs.~\cite{dim6decay}.}  These models make use of the
Froggatt-Nielsen mechanism \cite{froggatt} where a global ${\sf
  U(1)}_F$ symmetry is broken spontaneously by the vev of a gauge
singlet field $\Phi$ at a high scale. Then the Yukawa couplings arise
from the non-renormalizable operators,
\begin{align}
  h_{ij} = g_{ij} \left( \frac{\VEV{\Phi}}{\Lambda} \right)^{Q_i+Q_j}
  \ .
\end{align}
Here, $g_{ij}$ are couplings $\mathcal{O} \left( 1 \right)$ and $Q_i$
are the charges of the various fermions.  Particularly interesting is
the case with a `lopsided' family structure, where the chiral ${\sf
  U(1)}_F$ charges are different for the ${\sf 5}^*$ and ${\sf 10}$ of
the same family.

The difference between the 6D {\sf SO(10)} and 4D {\sf SU(5)} models
is most noticeable in the channel $p\to \mu^+ K^0$.  This is due to
the absence of second and third generation weak eigenstates in the
current-current operators and the vanishing (12)-component.  Hence the
decay $p\to \mu^+ K^0$ is doubly Cabibbo suppressed.  This effect is a
direct consequence of the localization of the `first generation' to
the Georgi-Glashow brane.

Let us now consider the general case, where the $\widetilde M^{d,e}$
are not degenerate and where $\mu_3$ and $\tilde\mu^{d,e}_3$ differ as
well.  From Eqn.~(\ref{eq:down-right}) we see that strange component
in $d^\cc_1$ does not vanish anymore, but it is smaller than the
bottom component.  We have studied several cases whose results agree
remarkably well.  As an illustration, consider the case where
$\tilde\mu^d_3=2\mu_3$ and $\tilde\mu^e_3=3\mu_3$, with non-degenerate
heavy masses \mbox{$\widetilde M^{d}_1 : \widetilde M^{d}_2 :
  \widetilde M^{d}_3 : \widetilde M^{d}_4 =
  \frac{1}{2}:\frac{1}{\sqrt{2}}:\frac{1}{\sqrt{2}}:1$} and
\mbox{$\widetilde M^{e}_1 : \widetilde M^{e}_2 : \widetilde M^{e}_3 :
  \widetilde M^{e}_4 = \frac{1}{2}:\frac{1}{\sqrt{2}}:1:\frac{1}{2}$}
(case II). The branching ratios are listed in Table~\ref{tb:result};
the differences between the two cases are indeed small.

The most striking difference between the 4D and 6D case is the decay
channel $p\to \mu^+ K^0$, which is suppressed by about two orders of
magnitude in the 6D model with respect to 4D models.  It is therefore
important to determine an upper limit for this channel in the 6D
model.  Varying the mass parameters in the range
$\widetilde{M}^{d,e}_\alpha/\widetilde{M}=0.1 - 1$ and
\mbox{$\tilde\mu_3^{d,e}/\mu_3=0.1-10$}, we find
\begin{align}
  \frac{\Gamma(p\to \mu^+ K^0)}{\Gamma(p\to e^+ \pi^0)} \lesssim 5\,\%
  \ ,
\end{align}
one order of magnitude smaller than in the four-dimensional case.
Note that in 5D orbifold GUTs, this channel is typically enhanced
because the first generation is located in the bulk
\cite{nomura01,hebecker02}.

Finally, we can derive the lower limit on the compactification scale
from the decay width of the dominant channel $p\to e^+ \pi^0$ yielding
$M_c \geq 9\times 10^{15}\,\text{GeV}$, roughly half of the 4D GUT
scale.


\section{Conclusion}

Supersymmetric GUTs provide a beautiful framework for theories beyond
the Standard Model.  The simplest model, minimal {\sf SU(5)}, is
inconsistent due to the failure of down quark and charged lepton
masses to unify.  The decay amplitude therefore depends strongly on
flavor mixing.  A consistent {\sf SU(5)} model requires additional
interactions, which are conveniently parameterized by higher
dimensional operators.  These operators can reduce the proton decay
rate by several orders of magnitude, hence proton decay does not rule
out consistent supersymmetric {\sf SU(5)} models.

Nevertheless, it is interesting to consider models where
dimension-five operators are absent; this is generically the case in
orbifold GUTs.  We discussed a 6D {\sf SO(10)} model, where the
position of the matter fields on the different branes leads to a
characteristic pattern of mass matrices.  The branching ratios differ
significantly from predictions of 4D GUTs, which can make it possible
to distinguish orbifold and four dimensional GUTs.  Furthermore,
proton decay restricts the compactification scale $M_c$ to be ${\cal
  O}\left(M_\gut^\text{4D}\right)$.

If proton decay is observed in the future, there will be only a few
events.  The dominance of the channel $p\to\bar\nu K^+$, however,
would strongly point at dimension-five operators, whereas $p\to
e^+\pi^0$ refers to dimension-six decay.  In the latter case, the
absence of the process $p\to\mu^+ K^0$ would disfavor the na{\"\i}ve
GUT model in four dimensions and test the idea that the different
generations are spatially separated.


\subsection*{Acknowledgements}

I am grateful to W. Buchm\"uller, L. Covi and D.  Emmanuel-Costa for
collaboration and I would like to thank them as well as A.
Brandenburg for a careful reading of the manuscript.


\end{document}